\documentclass[11pt,a4paper]{scrartcl}
\usepackage{adjustbox}

\newcommand{\sqrts}{\sqrt{s}}

\newcommand{\LumiInt}{\mathcal{L}_\mathrm{\tiny{int}}}
\newcommand{\alphas}{\alpha_{S}}

\newcommand{\uubar}{\ensuremath{\PQu\PAQu}\xspace}
\newcommand{\ddbar}{\ensuremath{\PQd\PAQd}\xspace}
\newcommand{\ssbar}{\ensuremath{\PQs\PAQs}\xspace}
\newcommand{\ccbar}{\ensuremath{\PQc\PAQc}\xspace}
\newcommand{\bbbar}{\ensuremath{\PQb\PAQb}\xspace}

\usepackage{ECFAreport}
\usepackage{amssymb}
\usepackage{amsmath}
\usepackage{comment}

\usepackage{ECFAreport_definitions}

\usepackage{xpatch}
\usepackage{placeins}

\DeclareOldFontCommand{\bf}{\normalfont\bfseries}{\mathbf}

\makeatletter
\xpatchcmd\@HepConStyle
 {\edef\@upcode{\updefault}}
 {\ifdefined\shapedefault\edef\@upcode{\shapedefault}\else\edef\@upcode{\updefault}\fi}
 {}{}
\makeatother

\RedeclareSectionCommand[
  beforeskip=1.\baselineskip,
  afterskip=.5\baselineskip]{subsection}
\RedeclareSectionCommand[
  beforeskip=1.\baselineskip,
  afterskip=.5\baselineskip]{subsubsection}

\DeclareCiteCommand{\citejournal}[\mkbibbrackets]
  {\usebibmacro{prenote}}
  {\usebibmacro{citeindex}%
   \printtext[bibhyperref]{\printfield{journaltitle}}%
   \iffieldundef{volume}
     {}%
     {\setunit{\addspace}%
     \printtext[bibhyperref]{\printfield{volume}}}%
   \setunit{\addspace}%
   \printtext[bibhyperref]{(\printdate)}%
   \iffieldundef{pages}
     {}
     {\setunit{\addspace}%
     \printtext[bibhyperref]{\printfield{pages}}%
     }%
     }
  {\multicitedelim}
  {\usebibmacro{postnote}}
  
\DeclareCiteCommand{\citesubmit}[\mkbibbrackets]
  {\usebibmacro{prenote}}
  {\usebibmacro{citeindex}%
   \printtext[bibhyperref]{\printfield{journaltitle}}%
   \setunit{\addspace}%
   \printtext[bibhyperref]{(\printdate)}}
  {\multicitedelim}
  {\usebibmacro{postnote}}
  
  \DeclareCiteCommand{\citeconf}[\mkbibbrackets]
  {\usebibmacro{prenote}}
  {\usebibmacro{citeindex}%
   \printtext[bibhyperref]{\printfield{howpublished}}%
   \setunit{\addspace}%
   \printtext[bibhyperref]{(\printdate)}}
  {\multicitedelim}
  {\usebibmacro{postnote}}

\usepackage{todonotes}
\usepackage{xpatch}
\makeatletter
\xpretocmd{\todo}{\@bsphack}{}{}
\xapptocmd{\todo}{\@esphack}{}{}
\makeatother

\usepackage{booktabs}
\usepackage{csquotes}

\DeclareTOCStyleEntry[beforeskip=.2cm]{section}{section}

\usepackage{tikz}
\usepackage{pgfplotstable}
\newcommand{\figmigration}[1]{
\pgfplotsset{
  /pgfplots/colormap={coldredux}{
    [1cm]
    rgb255(0cm)=(255,255,255)
    rgb255(2cm)=(0,192,255)
    rgb255(4cm)=(0,0,255)
    rgb255(6cm)=(0,0,0)
  }
}
\begin{tikzpicture}
  \begin{axis}[
    xlabel={$m_\mathrm{HFS}$\,\si{\GeV}},
    ylabel={$m(q\bar{q})$\,\si{\GeV}},
    xmin=0,
    xmax=100,
    ymin=0,
    ymax=100,
    minor tick num=4,
    xticklabel style={/pgf/number format/fixed},
    colorbar,
    scatter/use mapped color={%
      draw=mapped color,
      fill=mapped color
    },
            width=0.35\textwidth,
        height=0.35\textwidth
    ]
    \addplot[
      scatter,
      scatter src=explicit,
      only marks,
      mark=square*,
      mark size=6pt
    ] file{#1};
  \end{axis}
\end{tikzpicture}
}

\newcommand{\figcomparetwo}[1]{
\begin{tikzpicture}
\pgfplotsset{try min ticks=6}
\begin{axis}[
 ylabel=$\mathrm{d}\sigma_\mathrm{had}/\mathrm{d}m_\mathrm{HFS}\,\mathrm{pb}^{-1}\mathrm{GeV}^{-1}$,xlabel=$m_\mathrm{HFS}\,\mathrm{GeV}$,
    ymode=log,
    xmin=-1,xmax=99,
    ymin=0.1,ymax=100000.0,
    grid=none,
    grid style={line width=.1pt, draw=gray!10},
    major grid style={line width=.3pt,draw=gray!80},
    minor tick num=5,
    enlargelimits={abs=0.01},    
    ticklabel style={fill=none},
    xticklabel style={/pgf/number format/fixed, /pgf/number format/precision=0, /pgf/number format/fixed zerofill} ,
        xlabel style={at={(ticklabel* cs:1)},anchor=north west, xshift=-1.5cm, yshift=-0.0cm, font=\bf},
    ylabel style={at={(ticklabel* cs:1)},anchor=south west, xshift=-3.8cm, yshift=-0.6cm, font=\bf},
        legend style={draw=none, legend pos=north west,font={\bf},cells={align=left}},
        legend cell align={left},
        width=0.32\textwidth,
        height=0.35\textheight,
        yshift=0.1\textheight
    ] 
\addplot[mark=none,color=red, mark size=1.0pt,error bars/.cd,x dir=both,x explicit, error mark=none] table [x index=0, x error index=1,  y index=2,col sep=space] {#1};
\addlegendentry{$q\bar{q}$}
\addplot[mark=none,color=blue, mark size=1.0pt,error bars/.cd,x dir=both,x explicit, error mark=none] table [x index=0, x error index=1, y index=3,col sep=space] {#1};
\addlegendentry{$q\bar{q}\epem$}
\addplot[mark=none,color=green, mark size=1.0pt,error bars/.cd,x dir=both,x explicit, error mark=none] table [x index=0, x error index=1, y index=4,col sep=space] {#1};
\addlegendentry{$q\bar{q}\gamma$}
\addplot[mark=none,color=yellow, mark size=1.0pt,error bars/.cd,x dir=both,x explicit, error mark=none] table [x index=0, x error index=1, y index=5,col sep=space] {#1};
\addlegendentry{$q\bar{q}\mu^{+}\mu^{-}$}
\addplot[mark=none,color=brown, mark size=1.0pt,error bars/.cd,x dir=both,x explicit, error mark=none] table [x index=0, x error index=1, y index=6,col sep=space] {#1};
\addlegendentry{$q\bar{q}\nu\bar{\nu}$}
\addplot[mark=none,color=pink, mark size=1.0pt,error bars/.cd,x dir=both,x explicit, error mark=none] table [x index=0, x error index=1, y index=7,col sep=space] {#1};
\addlegendentry{$q\bar{q}\tau^{+}\tau^{-}$}
\addplot[mark=none,color=black, mark size=1.0pt,error bars/.cd,x dir=both,x explicit, error mark=none] table [x index=0, x error index=1, y index=8,col sep=space] {#1};
\addlegendentry{$\tau^{+}\tau^{-}$}
\addplot[mark=none,color=pink, mark size=1.0pt,error bars/.cd,x dir=both,x explicit, error mark=none] table [x index=0, x error index=1, y index=9,col sep=space] {#1};
\addlegendentry{$\tau^{+}\tau^{-}\gamma$}
\addplot[mark=none,color=orange, mark size=1.0pt,error bars/.cd,x dir=both,x explicit, error mark=none] table [x index=0, x error index=1, y expr=\thisrowno{10}+\thisrowno{11}+\thisrowno{12},col sep=space] {#1};
\addlegendentry{$\gamma\gamma\rightarrow q\bar{q}$}

\end{axis}
\end{tikzpicture}
}

\title{Physics case for low-$\sqrts$ QCD studies at FCC-ee}

\nolicence 

\addauthor{Andrii~Verbytskyi}{\institute{1}}
\addauthor{David~d'Enterria}{\institute{2}}
\addauthor{Pier~Francesco~Monni}{\institute{3}}
\addauthor{Peter~Skands}{\institute{4}}

\addinstitute{1}{Max-Planck Institut f\"ur Physik, Boltzmannstra{\ss}e 8
Garching 85748, Germany\\}
\addinstitute{2}{Experimental Physics Department, EP/FCC, CERN, CH-1211 Geneva, Switzerland\\}
\addinstitute{3}{Theoretical Physics Department, CERN, CH-1211 Geneva, Switzerland\\}
\addinstitute{4}{School of Physics \& Astronomy, Monash University, Clayton VIC-3800, Australia\\}

\abstract{\noindent
Measurements of hadronic final states in $e^{+}e^{-}$ collisions at centre-of-mass (CM) energies below the Z peak can notably extend the FCC-ee physics reach in terms of precision quantum chromodynamics (QCD) studies. Hadronic final states can be studied over a range of hadronic energies $\sqrt{s_\mathrm{had}} \approx 20\mbox{--}80\,\mathrm{GeV}$ by exploiting events with hard initial- and final-state QED radiation (ISR/FSR) during the high-luminosity Z-pole run, as well as in dedicated short (about one month long) $e^{+}e^{-}$ runs at CM energies $\sqrt{s} \approx 40\,\mathrm{GeV}$ and $60\,\mathrm{GeV}$. Using realistic estimates and fast detector simulations, we show that data samples of about $10^{9}$ hadronic events can be collected at the FCC-ee at each of the low-CM-energy points. Such datasets can be exploited in a variety of precision QCD measurements, including studies of light-, heavy-quark and gluon jet properties, hadronic event shapes, fragmentation functions, and nonperturbative dynamics. This will offer valuable insights into strong interaction physics, complementing data from nominal FCC-ee runs at higher center-of-mass energies, $\sqrt{s} \approx 91, 160, 240,$ and $365\,\mathrm{GeV}$.

\begin{center}
\vspace{3cm}
\large Back-up document to the ``FCC: QCD physics'' document submitted to ESPPU 2025
\end{center}
}

\begin{document}

\titlepage
\pagenumbering{arabic}\setcounter{page}{2}

\tableofcontents

\section{Introduction}

Measurements of hadronic final states in $\epem$ collisions offer a powerful avenue to study multiple aspects of the strong interaction in a very clean and controlled experimental and theoretical environment. These aspects include precise extractions of the quantum chromodynamics (QCD) coupling $\alphas$, the study of hadronic jet properties and jet observables, the testing and tuning of parton showers, extraction of parton-to-hadron fragmentation functions (FFs), and the study of nonperturbative phenomena such as hadronization and color reconnection~\cite{ALEPH:1996oqp,Bethke:2004bp, Kluth:2006bw,Proceedings:2017ocd,Accardi:2022oog}. 
These measurements offer crucial insights into QCD phenomena, enhancing the precision of Standard Model (SM) studies, including electroweak and Higgs measurements, and searches for physics beyond the SM. To date, detailed QCD studies in $\epem$ collisions have been carried out at low center-of-mass (CM) energies $\sqrt{s}\approx 3\mbox{--}10\,\GeV$, corresponding to (mini)jet energies $E_{j}\approx 1.5\mbox{--}5\,\GeV$, at the BEPC/BEPC-II collider (BES/BES-III experiments) and at B factories (BaBar, and Belle/Belle-II). 
Such studies have mostly focused on light- and heavy-quark hadron spectroscopy, as well as observables such as the $R(\sqrts) = \sigma(\epem\to\text{hadrons})/\sigma(\epem\to\mu^+\mu^-)$ ratio~\cite{Malaescu:2024wlo}. At CM energies of $\sqrts \approx 10\,\GeV$, the excellent statistics and particle identification (PID) capabilities of B-factories have enabled, in addition, determinations of parton-to-hadron FFs with high precision~\cite{Proceedings:2017ocd,Belle:2017caf,BaBar:2013jdt}, including for their transverse-momentum-dependent counterpart~\cite{Belle:2019ywy}, polarisations/helicities, and single-identified hadrons~\cite{Belle:2013lfg}, with further opportunities for future studies~\cite{Accardi:2022oog}. However, the range of validity of perturbative calculations at the CM energies reached by these machines is relatively limited, given that partons/jets have maximum energies of order $E_{j} \approx 5\,\GeV$, emphasizing the need for higher collision energies.

\begin{figure}[htpb!]
\adjincludegraphics[height=0.35\columnwidth,width=0.5\columnwidth]{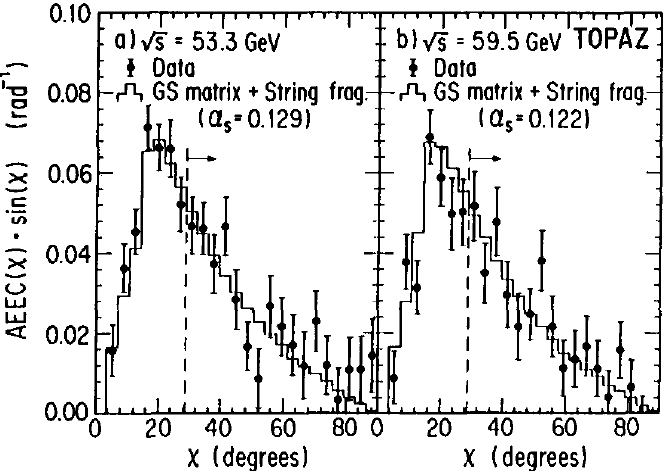}
\adjincludegraphics[height=0.345\textwidth,width=0.49\textwidth,clip,trim={0 0  {0.02\Width} {0.01\Height}}]{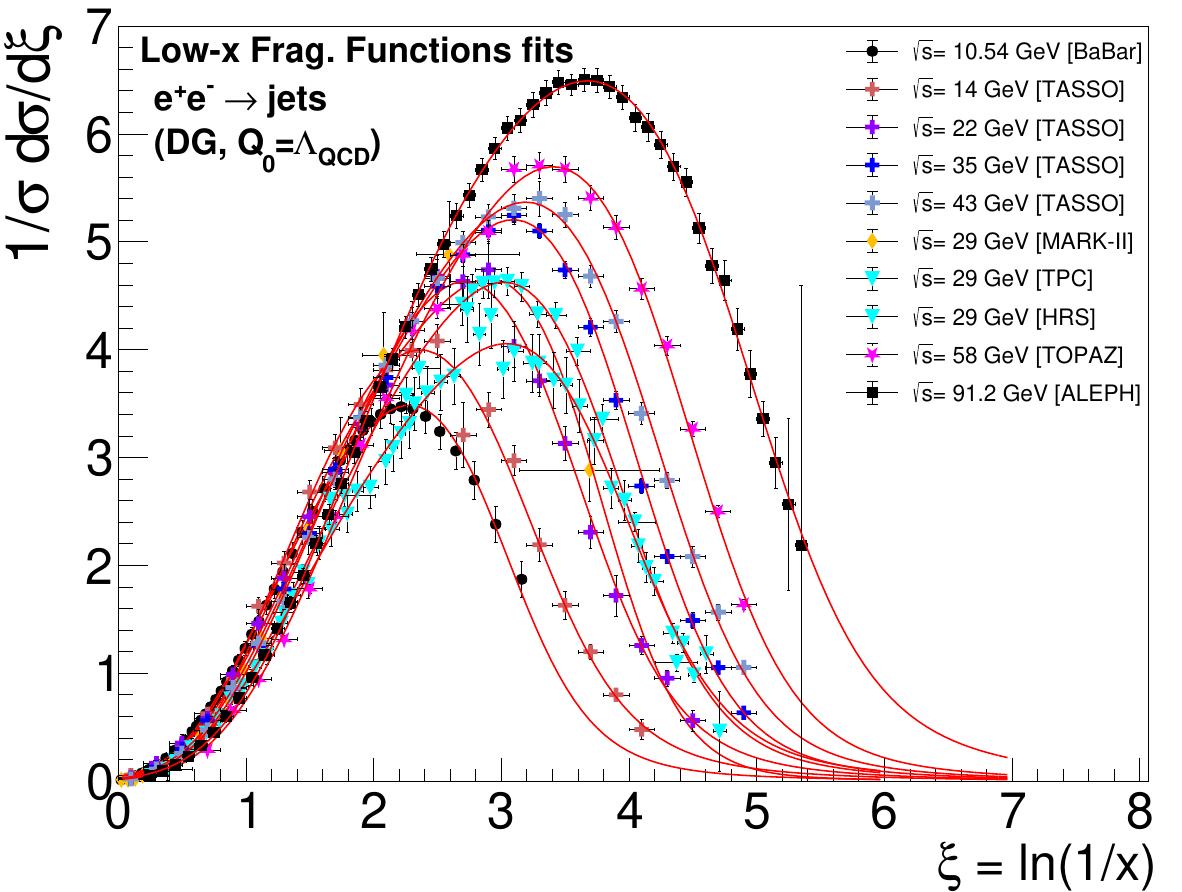}\\
\adjincludegraphics[height=0.35\textwidth,width=0.5\textwidth]{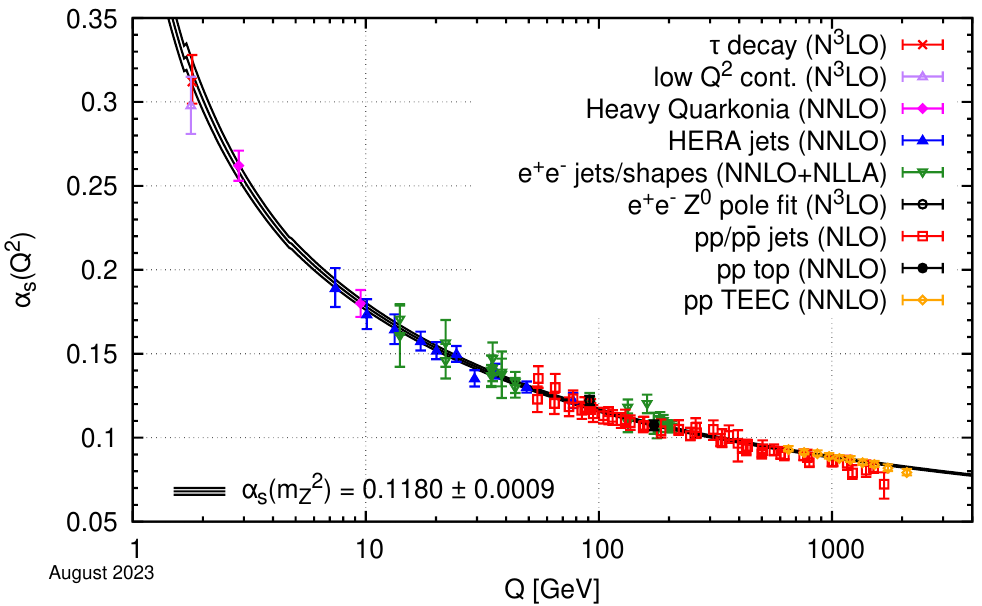}\adjincludegraphics[clip,trim={0 {0.02\Height} {0.054\Width} 0},height=0.345\textwidth,width=0.5\textwidth]{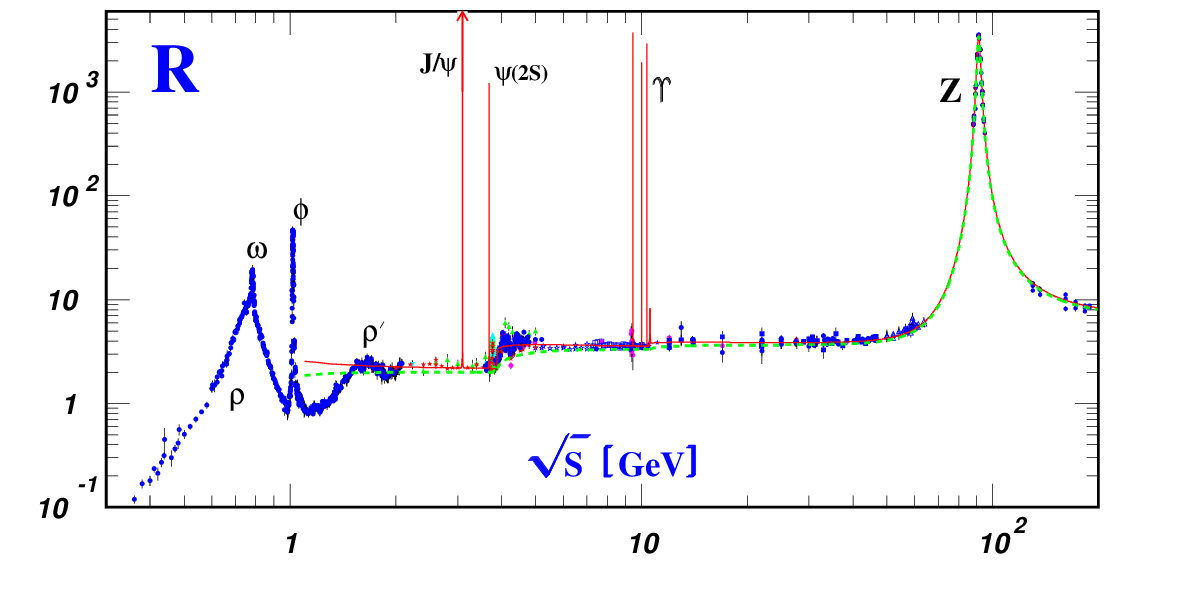}\\
\caption{Examples of existing QCD measurements using $\epem$ annihilation hadronic events over $\sqrt{s} \approx 10\mbox{--}80\,\GeV$, showing the size of their (typically large) experimental uncertainties: EEC asymmetry measured by TOPAZ~\protect\cite{TOPAZ:1989yod} (top left), fits of FFs at low hadron momenta~\protect\cite{TOPAZ:1994voc,TPCTwoGamma:1988yjh,Perez-Ramos:2013eba} (top right), $\alphas$ extractions from event shapes and jet rates~\protect\cite{ParticleDataGroup:2024cfk} (bottom left), and $R(\sqrt{s})$ ratio~\protect\cite{ParticleDataGroup:2024cfk} (bottom right). 
\label{fig:QCD_examples}}
\end{figure}

In the intermediate-energy range ($\sqrt{s} \approx 10\mbox{--}60\,\GeV$), QCD measurements exist from PETRA($12\mbox{--}47\,\GeV$) at DESY~\cite{Naroska:1986si,TASSO:1990cdg,Bethke:2009ehn}, PEP($29\,\GeV$) at SLAC~\cite{TPCTwoGamma:1984sji}, and TRISTAN($50\mbox{--}64\,\GeV$) at KEK~\cite{TOPAZ:1989yod,TOPAZ:1993vyh}, although with large statistical and systematic uncertainties compared to today's state-of-the-art. The combined number of $\epem$ annihilation events recorded by these experiments is one order-of-magnitude smaller than that collected at LEP-I energies. Moreover, such measurements were performed with detectors based on technologies that, while advanced at the time, have by now been superseded. Only a fraction of the recorded data was used for QCD physics analyses. Table~\ref{tab:old} provides estimates of the number of hadronic events recorded at TRISTAN, PETRA, and PEP. Examples of the experimental uncertainties for typical QCD measurements in this energy range are presented in Fig.~\ref{fig:QCD_examples} for energy-energy correlators (EEC) measured by TOPAZ at TRISTAN~\cite{TOPAZ:1989yod} (top left), $\alphas$ determinations from PETRA event shapes and jet rates~\cite{ParticleDataGroup:2024cfk} (top right), fits of low-$z$ parton-to-hadron FFs from TRISTAN and PEP (among other) data~\cite{TOPAZ:1994voc,TPCTwoGamma:1988yjh,Perez-Ramos:2013eba} (bottom left), and measurements of the $R$ ratio~\cite{Ezhela:2003pp} (bottom right).

The LEP/LEP-II experiments probed the highest $\epem$ CM energies to date, $\sqrt{s} \approx 88\mbox{--}210\,\GeV$ corresponding to jet energies $E_{j}\approx 44\mbox{--}105\,\GeV$, with an extended QCD program comprising the measurement of several observables (cf., \eg, the reviews~\cite{ALEPH:1996oqp,Bethke:2004bp, Kluth:2006bw} for detailed accounts). It is therefore of interest to understand to what extent the tenfold energy gap between the CM energies of B factories and the Z pole can be covered by measurements at a future $\epem$ facility.

\begin{table}[htbp!]
\tabcolsep=5mm
\caption{Estimates of the number of eligible hadronic events at TRISTAN, PETRA, and PEP. The numbers for PETRA are estimated by scaling by a factor of four the JADE results from Ref.~\cite{Schieck:2006tc}, \ie\ assuming the numbers for the MARK-J, TASSO, and CELLO experiments are reasonably similar. The numbers for TRISTAN were estimated scaling those from Ref.~\cite{TOPAZ:1989yod} to the total luminosity.\label{tab:old}}
\vspace{0.2cm}
\centering
\begin{tabular}{cccc}\hline
Accelerator  & CM energy range (\GeV)    & $\LumiInt$ ($\unit{pb}^{-1}$) & Number of eligible    \\
             &                         &                       & multihadron events  \\\hline
TRISTAN      &  $50\mbox{--}64$        &  $900$~\cite{Kuroda:1996zj}&  $1.1\times10^5$~\cite{TOPAZ:1989yod}     \\ 
PETRA        &  $12\mbox{--}47$        &  $760$~\cite{Schieck:2006tc}& $2\times10^5$~\cite{TASSO:1989kdk,Schieck:2006tc}      \\  
PEP          &  $29$           &  $315$~\cite{Allaby:1989dm}& $1.44\times10^5$~\cite{Allaby:1989dm}      \\\hline
\end{tabular}
\end{table}

The baseline program of the Future Circular Collider (FCC-ee)~\cite{FCC:2018byv,Bartmann:2928193} considers $\epem$ collisions at four interaction points (IPs) and at four nominal CM energies: 
$\sqrt{s} \approx 91, 160, 240, 360\,\GeV$, studies 
corresponding to a total number of $4\times 10^{12}$, $8\times 10^{8}$, $1.5\times 10^{8}$, $1.7\times10^{7}$ expected \textit{hadronic} final-state (HFS) events, respectively~\cite{FCC-PED-QCD_2025}. The Z-pole ($\sqrt{s} \approx 91\,\GeV$) and WW ($\sqrt{s} \approx 160\,\GeV$) runs at the FCC-ee will provide data samples that are many orders-of-magnitude larger than the corresponding $\mathcal{O}(17\times 10^6)$ 
and $\mathcal{O}(5\times 10^4)$ 
events collected at LEP at the same energies.
The FCC-ee multimillion-events data samples will enable detailed studies of heavy- and light-quark jets with maximum energies $E_\text{j}\approx 40\mbox{--}180\,\GeV$ from $\mathrm{Z^{(*)}, W, H} \to \qqbar$ decays, as well as of $\mathcal{O}(10^5)$ gluon jets with $E_\text{j}\approx 60\,\GeV$ from the $\mathrm{H}\to\Pg\Pg$ decays alone\footnote{The use of gluon data from Higgs decays for QCD studies must be carried out under the assumption that the $\mathrm{H}\to\Pg\Pg$ partial width is SM-like or, alternatively, ensuring that the analysis considered is agnostic to the details of the loop-induced $\mathrm{H}\to\Pg\Pg$ decay.}. Although an even larger sample of lower-energy gluon jets can be also studied at the Z-pole run (via $\PZ\to \bbbar\,\Pg$ with double-tagged b jets), the range of $E_{j}\approx 5\mbox{--}40\,\GeV$ jet energies, as well as the overall properties of hadronic final states in events with total invariant mass $\sqrt{s_\text{had}}\approx 20\mbox{--}80\,\GeV$, will remain poorly explored. This note presents the case for QCD studies in $\epem$ collisions below the Z-pole at the FCC-ee, and assesses the means to perform the potential collection of such sets of data.

From a theoretical viewpoint, the information encoded in data collected at different CM energies can be used to disentangle more effectively the effect of hadronization corrections to collider observables in ways that cannot be achieved with data at a single CM energy. 
In order to make this point more concrete, let us focus on classes of key observables such as global event shapes or jet rates, generically denoted by the dimensionless variable $v$, measured in $\epem \to \PZ^*/\gamma^*\to$~jets. In the perturbative regime, $v \gg \Lambda_\text{QCD}/\sqrt{s_\text{had}}$ where $\Lambda_\text{QCD}$ is a typical hadronic scale and $\sqrt{s_\text{had}}$ is the scale of the hadronic final state (which can be smaller than $\sqrt{s}$ in the presence of initial- and/or final-state QED radiation), the differential cross section $\mathrm{d}\sigma\equiv \mathrm{d}\sigma/d v$ can be schematically parametrized as,
\begin{equation}
\mathrm{d}\sigma \sim \mathrm{d}\sigma^\text{{(P)}} + \mathrm{d}\sigma^\text{{(NP)}}\,.
\end{equation}
Here, $\mathrm{d}\sigma^\text{{(P)}}$ denotes the perturbative part of the theoretical prediction, which depends on the observable $v$ as well as on ratios of scales characteristic of the scattering process, such as $\sqrt{s_\text{had}}$ and heavy-quark masses $m_\mathrm{Q}$ (\eg, that of the charm or bottom quarks).
The quantity $\mathrm{d}\sigma^\text{{(NP)}}$ denotes the nonperturbative correction to the cross section due to hadronization that, in the limit $v \gg \Lambda_\text{QCD}/\sqrt{s_\text{had}}$, can be expanded in powers of $\Lambda_\text{QCD}/\sqrt{s_\text{had}}$, typically starting with a linear term, \ie 
\begin{equation}
\mathrm{d}\sigma^\text{{(NP)}}\sim \Omega(v)\,\left(\frac{\Lambda_\text{ QCD}}{\sqrt{s_\text{had}}}\right) + \mathcal{O}\left(\frac{\Lambda^2_\text{ QCD}}{s_\text{{had}}}\right)\,,
\end{equation}
with a coefficient $\Omega(v)$ that depends on the observable itself.
Due to this scaling, measurements at different hadronic energies help disentangle these effects from the perturbative contribution to an observable's theoretical prediction, which can be exploited to gain valuable insight into their field-theoretical description.
Moreover, low-energy measurements are crucial to improve models of hadronization dynamics and their tuning in Monte Carlo (MC) event generators, which will be essential in analyzing high-precision measurements for Z and H decays, as well as for simulations at the WW and $\ttbar$ thresholds.
A second aspect that will be possible to study with runs at different CM energies is the impact of $m_\mathrm{Q}$ effects, \eg\ the \textit{dead cone}~\cite{Dokshitzer:1991fd}, on collider observables, which commonly scale as $\mathcal{O}(m_\mathrm{Q}^2/s_\text{{had}})$ in $\mathrm{d}\sigma^\text{{(P)}}$.
One can use these data to test and improve the theoretical description of charm and bottom quarks and their hadronization in theoretical calculations as well as in event generators, which will be instrumental in the study of the properties of heavy-quark jets.
Finally, data at different CM energies allows for alternative extractions of the strong coupling $\alphas$ using \eg\ event-shape observables. Although these extractions would arguably not be competitive with the one foreseen at the Z-boson resonance via inclusive Z decays into hadrons~\cite{dEnterria:2020cpv}, they can be used to shed new light on a long-standing discrepancy between different $\alphas$ fits from event shapes (we refer the reader to the discussion in Ref.~\cite{ParticleDataGroup:2024cfk}).

In summary, a comprehensive set of QCD measurements performed at the FCC-ee across a broad range of CM energies will provide unique insights into nonperturbative phenomena and enhance the understanding of heavy-quark mass effects on a range of key collider observables. 
A machine such as FCC-ee offers two possible strategies to access lower hadronic energies. The simplest and most straightforward solution is to make use of events with initial- or final-state (ISR/FSR) hard QED radiation during the $\sqrt{s} =m_{\PZ}$ run, and exploiting a fraction of the enormous, $\mathcal{O}(6\times10^{12})$ events, sample to be collected (Section~\ref{sec:ISR_FSR}). ISR/FSR events are characterized by the emission of one (or more) hard photon(s), mostly collinear to the $\Pepm$ beam, thereby reducing the effective CM energy $\sqrt{s^\prime}$ of the colliding $\epem$ pair, and boosting the hadronic final state relative to the LAB frame. A complementary option is to set up and perform dedicated runs at individual $\epem$ collision energies below $m_{\PZ}$. Given the very large instantaneous luminosities available above the injection beam energy of $E_\mathrm{beam} = 20\,\GeV$ from the top-up booster, such runs could presumably be done within a short running time, about one month at each energy point, as estimated in Section~\ref{sec:dedicated_runs}. 

\section{Low-$\sqrt{s}$ hadronic data samples from ISR/FSR events at the Z pole run}
\label{sec:ISR_FSR}

The LEP collider never operated $\epem$ collisions significantly below $\sqrt{s} = m_{\PZ}$, but all LEP experiments collected ISR/FSR events, characterized by the emission of one or more photons that lower energy flowing into the hadronic final state, and exploited them for QCD physics analyses. Based on the LEP experience, we first estimate the expected size of the ISR/FSR data samples that can be collected at FCC-ee, and we then study the purity and efficiency of the HFS event selection criteria by means of a fast simulation based on the IDEA detector concept~\cite{Antonello:2020tzq}.

\subsection{Estimates of the size of ISR/FSR event samples}

As a reference, we use the L3 study of Ref.~\cite{L3:2004cdh} which featured $\LumiInt \approx 140$~pb$^{-1}$ integrated over a large span of hadronic CM energies. With this setup, it is possible to estimate the number of HFS events suitable for QCD analyses to be collected at FCC-ee at reduced CM energies, assuming similar ISR/FSR event selection criteria and scaling the corresponding sample sizes to the $\mathcal{O}(10^6)$ times larger luminosities. The selection criteria applied in Ref.~\cite{L3:2004cdh} were meant to collect a data sample for the measurement of event shape observables, and is very similar to the analyses carried out by other LEP experiments, \eg, by OPAL~\cite{OPAL:2007haf} and DELPHI~\cite{DELPHI:2003yqh}. The typical ISR/FSR data samples collected by L3, as well as the corresponding number of events estimated for FCC-ee at the Z pole, are listed in Table~\ref{tab:l3}. Here one can see that hadronic samples with $\mathcal{O}(10^9)$ events can be collected at each reduced CM energy $\sqrt{s^\prime}$, derived from the energy of the ISR/FSR photon $E_{\gamma}$ as $\sqrt{s^\prime}=2\times E_\text{beam} \sqrt{1-\frac{E_{\gamma}}{E_\text{beam}}}$ (in the CM frame).
The selected event samples at the FCC-ee would be $10^3$ larger than those collected by all previous experiments in $\epem$ collisions at dedicated intermediate CM energies (Table~\ref{tab:old}).

\begin{table}[htbp]
\caption{Characteristics of the hadronic data samples collected via ISR/FSR events by the L3 experiment~\protect\cite{L3:2004cdh}, and the estimated number of events that could be obtained at FCC-ee~(4 IPs combined) with the expected $\LumiInt~=~205$~ab$^{-1}$ at the Z pole (rightmost column). 
\label{tab:l3}}
\vspace{0.2cm}
\centering
\resizebox{\textwidth}{!}{%
\begin{tabular}{ccc@{.}lr@{.}lccr|c}\hline
Type  &$\sqrt{s^\prime}$ (\GeV)  &\multicolumn{2}{c}{$\langle\sqrt{s^\prime}\rangle$ (\GeV)}&\multicolumn{2}{c}{$\LumiInt$~(pb$^{-1}$)}& Sel. Eff. (\%) & Purity (\%) & \# Sel. Evts (L3)& \# Sel. Evts (FCC-ee)\\\hline
            & 30--50     &    41&4 &\phantom{00} 142&4 &  48.3     & 68.4   &    1247    &$1.8\times 10^{9}$  \\
Reduced     & 50--60     &    55&3           & 142&4      &  41.0     & 78.0   &    1047 &$1.4\times 10^{9}$\\
CM          & 60--70     &    65&4           & 142&4      &  35.2     & 86.0   &    1575 &$2.2\times 10^{9}$ \\
Energy      & 70--80     &    75&7           & 142&4      &  29.9     & 89.0   &    2938 &$4.2\times 10^{9}$ \\
            & 80--84     &    82&3           & 142&4      &  27.4     & 90.5   &    2091 &$3.0\times 10^{9}$ \\
            & 84--86     &    85&1           & 142&4      &  27.5     & 87.0   &    1607 &$2.2\times 10^{9}$\\  \hline
Z pole      & 91.2       &    91&2           &   8&3      &  98.5     & 99.8   &  248\,100 &$4.5\times 10^{12}$ \\\hline  
\end{tabular}
}
\end{table}

\begin{figure}[htpb!]
\centering
\includegraphics[width=0.49\columnwidth]{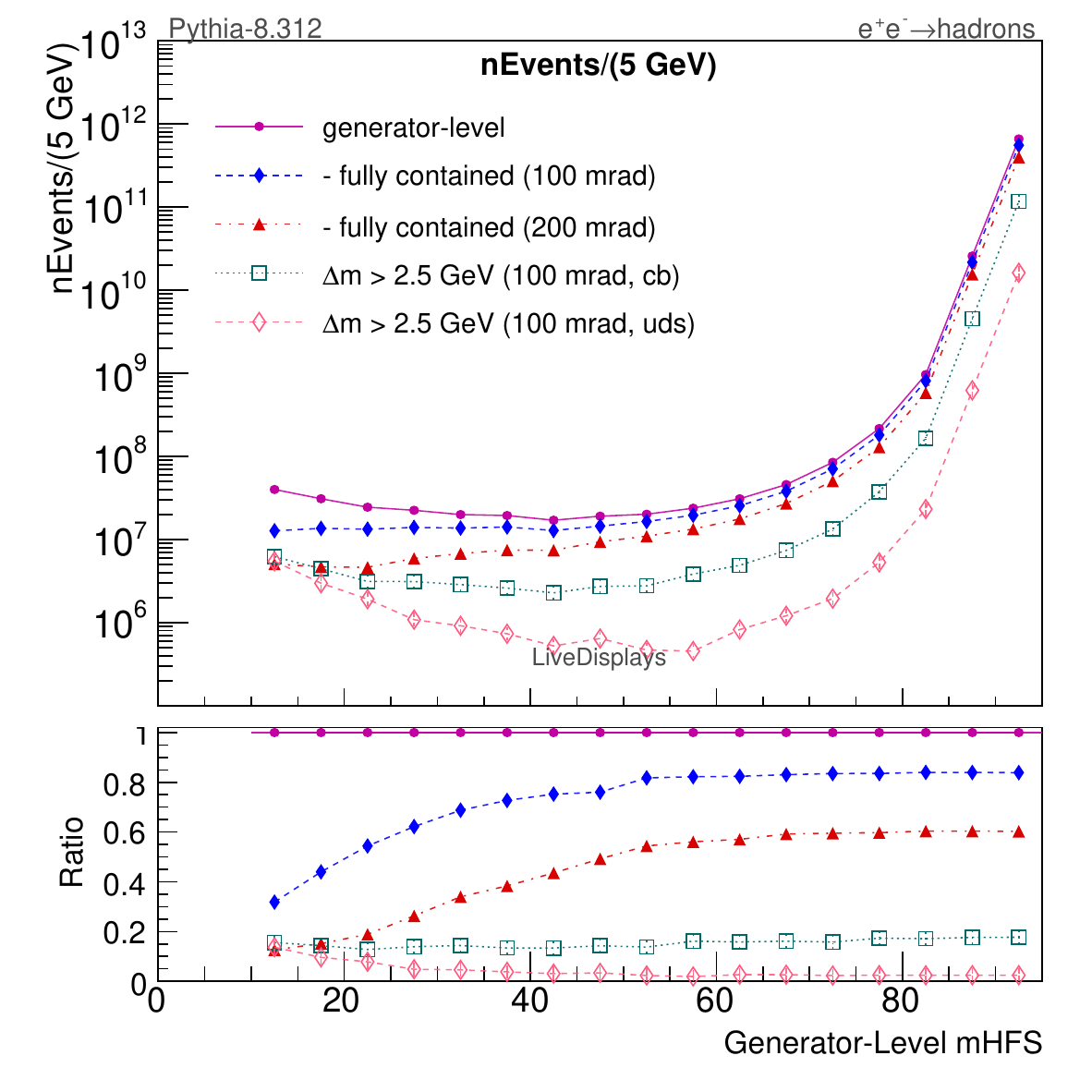}
\includegraphics[width=0.49\columnwidth]{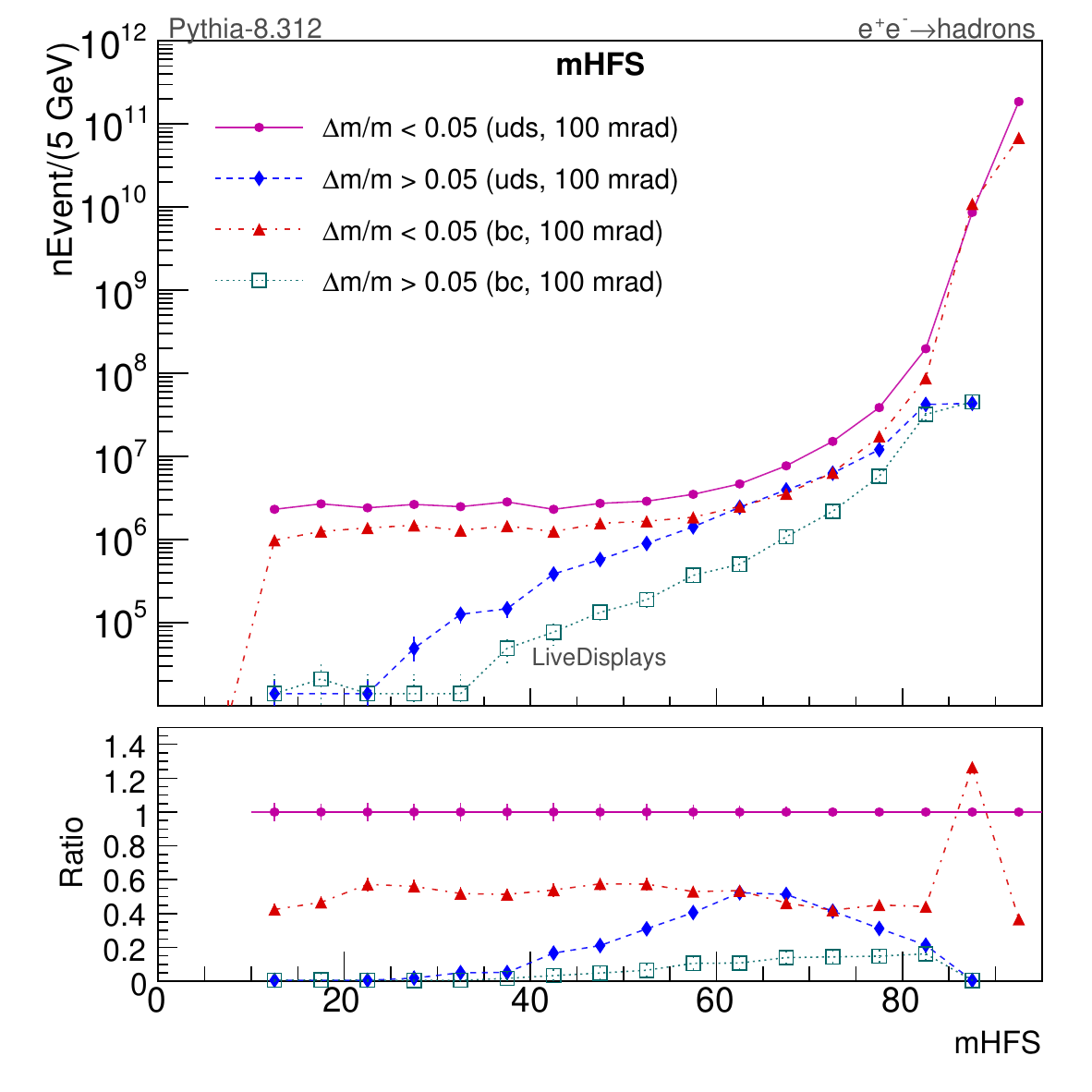}
\caption{Number of ISR/FSR events as a function of the HFS mass (in 5-$\GeV$ bins) corresponding to a run with $10^{12}$ $\epem \to \qqbar$ events at $\sqrt{s} = 91.2\,\GeV$, generated with PYTHIA~8~\protect\cite{Bierlich:2022pfr}. The left figure shows the number of hadronic events fully contained within a detector with geometric acceptance down to $\theta_\text{min} = 100$~mrad (FCC-ee-like) and 200~mrad (LEP-like) as a function of the generator-level hadronic CM energy. The same plot also shows the number of hadronic events for well-reconstructed light-quark (u,\,d,\,s) and heavy-quark (c,\,b) jets for a coverage down to $\theta_\text{min} = 100$~mrad. The right figure shows the number of $\epem\to \uubar,\ddbar,\ssbar$ and $\epem\to\ccbar,\bbbar$ events as a function of their visible hadronic invariant mass, for those well ($\Delta m/m=|\sqrt{s_\text{had,true}} - m_\mathrm{HFS}|/\sqrt{s_\text{had,true}} < 5\%$) and badly ($\Delta m/m=|\sqrt{s_\text{had,true}} - m_\mathrm{HFS}|/\sqrt{s_\text{had,true}} > 5\%$) reconstructed for a detector with acceptance down to $\theta_\text{min} = 100$~mrad.
\label{fig:m_had}}
\end{figure}

An additional advantage of the FCC-ee with respect to LEP experiments is their comparatively enlarged geometric detector acceptance. The FCC-ee tracker (calorimeter) is expected to reach polar angles down to $\theta \approx 100~(100)$~mrad, instead of $\theta \approx 250~(200)$~mrad at LEP. 
Such an extended acceptance (covering the $|\cos\theta|\lesssim 0.99$, instead of $|\cos\theta|\lesssim 0.96$, range) allows the full reconstruction of hadronic ISR/FSR events down to $\sqrt{s_\text{had}} \approx 10\mbox{--}20\,\GeV$, instead of the minimum $\sqrt{s_\text{had}} \approx 30\,\GeV$ reached at LEP. 
This is shown in Fig.~\ref{fig:m_had} (left) by comparing the number of events (the blue rhombus and red triangle curves) as a function of the reconstructible visible mass distribution, $m_\mathrm{HFS}$, \ie, the invariant mass of all non-neutrino particles within the acceptance, defined as\footnote{This quantity is meant to be a proxy for the CM energy of the hadronic system. For events with an identified FSR photon, the photon is excluded from the sum.} 
\begin{equation}
m_ \text{HFS} = \sqrt{\left(\Sigma_{i}^{N} E_i\right)^2 - \left(\Sigma_{i}^{N}\vec{P}_i\right)^2},\; \mbox{ with } N = \mbox{all visible particles in the detector}.
\end{equation}
The curves are obtained at the generator-level with PYTHIA\,8~\cite{Bierlich:2022pfr} without detector effects, except by requiring full containment of the hadronic events within different acceptance ranges. As an additional piece of information, 
Fig.~\ref{fig:m_had} (right) shows the number of ISR/FSR events for $\epem\to \uubar,\ddbar,\ssbar$ and $\epem\to\ccbar,\bbbar$ final states, separately.
This plot shows the expected number of light- and heavy-quark events, divided into those that have $\Delta m/m=|\sqrt{s_\text{had,true}} - m_\mathrm{HFS}|/\sqrt{s_\text{had,true}} < 5\%$ (well reconstructed signal, $m_\mathrm{HFS}$ is close to the true CM energy of the HFS) and those that have $\Delta m/m=|\sqrt{s_\text{had,true}} - m_\mathrm{HFS}|/\sqrt{s_\text{had,true}} > 5\%$ (badly reconstructed, background-like).

\subsection{Simulated ISR/FSR event selection studies}

From the past LEP studies~\cite{L3:2004cdh,OPAL:2007haf,DELPHI:2003yqh}, it is also possible to identify the challenges one faces in the analysis of the ISR/FSR events, such as the presence of $\mathcal{O}(1\%)$ background events from $\epem\to\tau^{+}\tau^{-}(\gamma)$ processes, and the $\mathcal{O}(10\%)$ contamination from photons coming from hadron decays and from misidentified photons (e.g. a $\pi^0$ misidentified as a photon).
Also, for ISR/FSR events above $\sqrts' \approx 40\,\GeV$, one increasingly faces spillover from escaping particles and misreconstruction of events at or near the Z peak, where the cross section is vastly higher. The ISR/FSR event selection strategy and the impact of the backgrounds are assessed next using FCC-ee fast simulations.

The processes with the largest cross section in $\epem$ collisions at $\sqrts\approx m_{\PZ}$ are $\gamma\gamma\to\,$hadrons (including both direct and resolved processes\footnote{Here, \textit{direct} refers to configurations in which the photon interacts as a pointlike particle, while \textit{resolved} refers to configurations in which it effectively interacts as a hadronic state, typically treated in the framework of Vector Meson Dominance (VMD).}, albeit all peaking at low HFS masses) and the production of two or four fermions in the final state accompanied by ISR/FSR photons (including also FSR gluon radiation from quarks). Several of these processes will have a boosted $\qqbar$ pair with $m(\qqbar)<2\times E_\text{beam}$ in the final state, and will be observed in the detector as multihadron events. To obtain a realistic picture of the HFS in the detector and be able to reconstruct the centre of mass of the boosted hadronic system, it is necessary to simulate all background processes and apply realistic event selection procedures.

The ISR/FSR analyses from previous LEP studies~\cite{L3:2004cdh,DELPHI:2003yqh,OPAL:2007haf} can be applied to the simulated FCC-ee Z-pole samples to select events with quarks and gluons in the final state with well-reconstructed kinematics in three different scenarios: $a)$ assuming the production of a single high-energy ISR/FSR photon, $b)$ assuming the production of a ISR/FSR photon collinear to the beam, and $c)$ assuming negligible ISR/FSR emission. The requirements to select each type of such events are as follows:
\begin{enumerate}
\item[$a)$] Enough visible hadrons in the final state within the detector acceptance\footnote{\label{multi}E.g., at least five tracks or calorimeter objects.}, requiring that the total visible energy $E_\mathrm{vis}$ does not deviate much from $2\times E_\text{beam}$. In addition, a well isolated high-energy\footnote{Eg., at least $E_\gamma = 10\,\GeV$.} photon with energy $E_{\gamma}$ is registered in the detector. The HFS without the photon is clustered into two jets which should satisfy the triangle condition, see Eq.~(3) of Ref.~\cite{DELPHI:2003yqh} for details\footnote{The photon energy can be also estimated clustering the remaining HFS into two jets $j_1$ and $j_2$ and using the sinus theorem: 
$$
E_{\gamma,\text{triangle}} = 2\times E_\text{beam}\times \frac{|\sin{j_1\wedge j_2}|}{|\sin{j_1\wedge j_2}| +|\sin{j_1\wedge \gamma}| +|\sin{j_2\wedge \gamma}| }. 
$$ $E_\mathrm{\gamma}$ should lie in the $[E_{\gamma,\text{triangle}}-10\,\GeV,E_{\gamma,\text{triangle}}+5\,\GeV]$ interval and the photon should be isolated from the jets such that $min( j_1\wedge \gamma,j_2\wedge \gamma)>0.5$.
}. This selection aims at selecting wide-angle high-energy ISR/FSR events and reconstruct their kinematics correctly.

\item[$b)$] Enough visible hadrons in the final state within the detector acceptance\footref{multi}, requiring that the total visible energy $E_{vis}$ does not deviate much from $2\times E_\text{beam}-|P_{\mathrm{vis},z}|$, where
$P_{\mathrm{vis},z}$ is the longitudinal component of the total visible momenta. The latter condition implies an existence of a single ISR photon radiated parallel to the beam and not registered in the detector, which is almost completely responsible for the momentum imbalance in the event\footnote{Therefore the requirement ($\vec{P}_\mathrm{vis}\wedge beam < 3^{\circ}$ or $\vec{P}_\mathrm{vis}\wedge beam >177^{\circ}$) is imposed.}. The events should also fail the criterion a).
This selection is designed to select events with ISR/FSR photons collinear to the beam direction and reconstruct the kinematics of these events correctly.

\item[$c)$] Enough visible hadrons in the final state within the detector acceptance\footref{multi}, requiring that the total visible energy $E_\mathrm{vis}$ does not deviate much\footnote{\label{endiff}E.g., less than 5\,\GeV.} from $2\times E_\text{beam}$,
and that the thrust vector direction is contained within the detector acceptance range\footnote{$|\cos{\theta_{T}}|<0.9$.}. The events should also fail the criterion a).
This selection is aimed at selecting events without significant ISR/FSR and reconstruct the kinematics of these events correctly.
\end{enumerate}

\begin{figure}[htpb!]
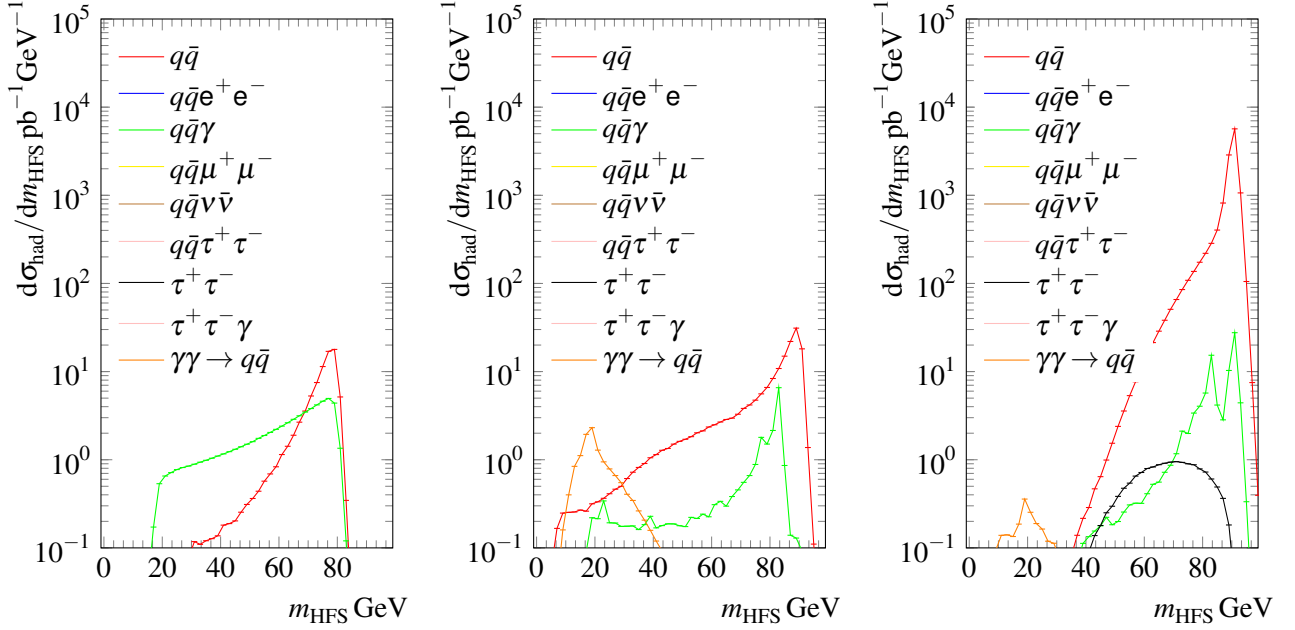

\centering\boldmath
\figcomparetwo{data3/SFSR.dat}\figcomparetwo{data3/SISR.dat}\figcomparetwo{data3/SBorn.dat}
\caption{
Distribution of the visible HFS invariant mass for multiple processes in $\epem$ collisions at $\sqrt{s} = 91.2\,\GeV$ that pass the three kinematic selection criteria $a),\,b),\,c)$, outlined in the text. The ISR/FSR photon is excluded from the HFS mass calculation. All final states but $q\bar{q}$, $q\bar{q}\gamma$, and $\tau^+\tau^-$ are strongly suppressed by the selection requirements. The full visible signal in the detector will be the sum of the displayed processes. Left: Events passing selection $a)$, with large purity for $\qqbar\gamma$ samples. 
Center: Events passing selection $b)$, with high purity for $q\bar{q}$ samples with collinear radiation.
Right: Events passing selection $c)$, with large purity for $q\bar{q}$ samples with negligible ISR/FSR emission.
\label{fig:delphes}}
\end{figure}

The three event-selection criteria outlined above were applied to the simulated FCC-ee data samples at the hadron-level as well as smeared to take into account detector effects. The ISR/FSR signal and background events were generated with the SHERPA~3.0.1 MC event generator~\protect\cite{Sherpa:2019gpd,Bothmann:2024wqs}, and the smearing was performed using the DELPHES card~\protect\cite{deFavereau:2013fsa} corresponding to the IDEA detector~\cite{Antonello:2020tzq,DOnofrio:2024gqh}.
The distributions of the visible HFS invariant mass for the events that passed the selections at the smeared level are shown in Fig.~\ref{fig:delphes}. This figure indicates that it is relatively easy to select $\epem\to \mathrm{hadrons}$ events of the type $c)$, with $m_\mathrm{HFS}\approx91\,\GeV$ with $\approx 100\%$ purity, \ie\ coming from $\epem\to \qqbar$ process with negligible ISR/FSR. On the other hand, the selection $a)$ will provide samples of $\epem\to \mathrm{hadrons} + \gamma_\text{ISR/FSR}$ events with purity $\approx 90\%$ for $20<m_\mathrm{HFS}<55\,\GeV$, and the selection $b)$ can be used to obtain samples of $\epem\to \mathrm{hadrons} + \gamma_\text{ISR/FSR}$ events with $\approx 90\%$ purity for $30<m_\mathrm{HFS}<70\,\GeV$. These findings are quite consistent with the L3 experiment results (Table~\ref{tab:l3}).

\begin{figure}[htpb!]
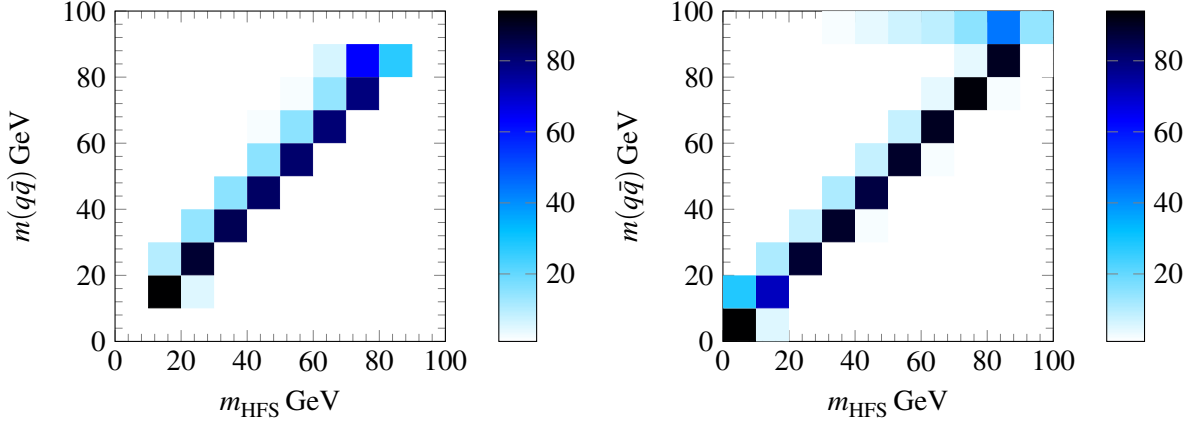

\centering\boldmath
\figmigration{data/d2SFSR.dat}\figmigration{data/d2SISR.dat}
\caption{
Correlation between the true mass ($m(\qqbar)$) and the HFS mass reconstructed at the detector level ($m_\mathrm{HFS}$) for $\epem\to \mathrm{hadrons} + \gamma_\text{FSR}$ events passing the selection $a)$ (left),
and for $\epem\to \mathrm{hadrons}$ ISR events passing the selection $b)$ (right). The values are normalized across the $x$ axis and the color coding scale is given in \%.
\label{fig:migration}
}
\end{figure}

Figure~\ref{fig:migration} compares the correlation between the reconstructed ($m_\mathrm{HFS}$) and the \textit{true} (parton level) HFS mass ($m(\qqbar)$) for the selected events. These preliminary studies indicate that with a 5-$\GeV$ bin width\footnote{A finer than 5-$\GeV$ binning could potentially be obtained with more sophisticated methods of event selection and kinematic reconstruction (
e.g.\ more sophisticated combinations of the kinematics information on the photon, HFS and beams). 
Further refined studies will be necessary to explore more in detail the reach of ISR/FSR events.} for $m(\qqbar)$, between $50\%$ and $90\%$ of the events have $m_\mathrm{HFS}$ reconstructed within the \textit{true} bin. The resolution of reconstructed HFS invariant-mass is sufficient for many studies, but it is intrinsically limited by fluctuations of energy escaping detection (\eg, taken away by neutrinos). Therefore, for certain precision studies (\eg, $\alphas$ extractions from event shapes, where the ISR-based studies~\cite{Abbiendi:2007aa} have larger uncertainties than those from dedicated low-$\sqrts$ runs~\cite{Bethke:2008hf}), the data from dedicated low-energy runs (discussed next) may be strongly preferred over that measured in ISR/FSR events.

\section{Potential dedicated low-$\sqrt{s}$ runs at FCC-ee}
\label{sec:dedicated_runs}

An alternative option to study HFS in $\epem$ collisions below $\sqrt{s} = m_{\PZ}$ is to perform dedicated FCC-ee runs at low CM energies. The lowest nominal CM energy at FCC-ee can be achieved in collisions at the design initial energy ($E_\mathrm{beam} = 20\,\GeV$) of the $\Pepm$ beams injected from the top-up booster~\cite{FCC:2018evy}, i.e.\ $\sqrt{s} = 2\times E_\mathrm{beam} = 40\,\GeV$. Table~\ref{tab:sa} lists the cross section for the $\epem\to\qqbar$ process (ISR included) over $\sqrt{s} = 40\mbox{--}80\,\GeV$, computed with SHERPA~3.0.1. In the previous section it was shown that data sets with $\mathcal{O}(10^9)$ hadronic events can be obtained via ISR/FSR studies at the Z pole, and we now want to estimate how many days ($t = 0.86\times10^5$s of physics) it would take to collect the same number of HFS events in dedicated low-$\sqrt{s}$ runs. The simplest 
scenario is to assume that the luminosity follows the standard Lorentz $\gamma_\mathrm{L}$-dependence driven by the beam energies\footnote{Such a dependence is approximately confirmed with the simulated collider parameters of Table~\ref{tab:FCCee_low_sqrts}.}, namely $\mathcal{L}\propto \sqrt{s}$. The number of days required to collect $\mathcal{O}(10^9)$ HFS events in such a scenario are listed in Table~\ref{tab:sa}. One would need about one month of running time, including 1--3 weeks of operation (depending on the $\sqrt{s}$ and actual $\LumiInt$ achieved) plus 1-week of additional beam-setup time~\cite{Frank2024}, for each single CM-energy point.

\begin{table}[htbp]
\tabcolsep=4mm
\centering
\caption{Integrated luminosity ($\LumiInt$) and time (days) needed to collect $10^9$ hadronic events (4 IPs combined) in dedicated FCC-ee runs at low CM energies $\sqrt{s} = 40\mbox{--}80\,\GeV$ (with corresponding cross section, including ISR, indicated), assuming that the instantaneous luminosity follows a $\mathcal{L}\propto \sqrt{s}$ scaling.}
\vspace{0.2cm}
\label{tab:sa}

\begin{tabular}{cccc}
\hline
$\sqrt{s}$ (\GeV) & $\sigma(\epem\to\qqbar)$ (ISR incl.) & \multicolumn{2}{c}{Integ.\ lumi and time needed to collect $10^9$ HFS events }\\
                &  (pb) &  ~~~~~~~~$\LumiInt$ (ab$^{-1}$) & time \\\hline
80              & 413  &  2.5 &      6 days    \\ 
70              & 182  &  5.5 &      14 days   \\  
60              & 162  &  6.2 &      20 days   \\  
50              & 200  &  5.0 &      18 days   \\  
40              & 289  &  3.5 &      23 days   \\\hline  
\end{tabular}
\end{table}

The technical feasibility of operating the FCC-ee at CM energies below the Z pole has not been studied, but parametric simulations of the accelerator settings have been kindly provided by K.~Oide~\cite{Katsunobo}, starting off from the beam settings for the Z-pole run, without further modifications of the machine. The results of such simulations are shown in Table~\ref{tab:FCCee_low_sqrts} for runs at $\sqrt{s} = 40, 60\,\GeV$. The last row of the table lists the achievable luminosities that quantitatively support the $\mathcal{L}\propto \sqrt{s}$ scenario of Table~\ref{tab:sa}, and therefore the possibility to obtain the desired $\mathcal{O}(10^9)$ HFS data samples within about 1-month of operation. Naturally, such a possibility should be investigated with dedicated realistic studies taking into account all relevant machine and detector aspects.

\begin{table}[h!]
\caption{FCC-ee collider parameters for the default Z-pole run (45.6-$\GeV$ column) and for potential runs at $\sqrt{s} = 40, 60\,\GeV$ (two rightmost columns)~\cite{Katsunobo}.}
\vspace{-0.5cm}
\centering
\begin{tabular}{c}\\
\raisebox{-\totalheight}{\includegraphics[width=0.95\textwidth]{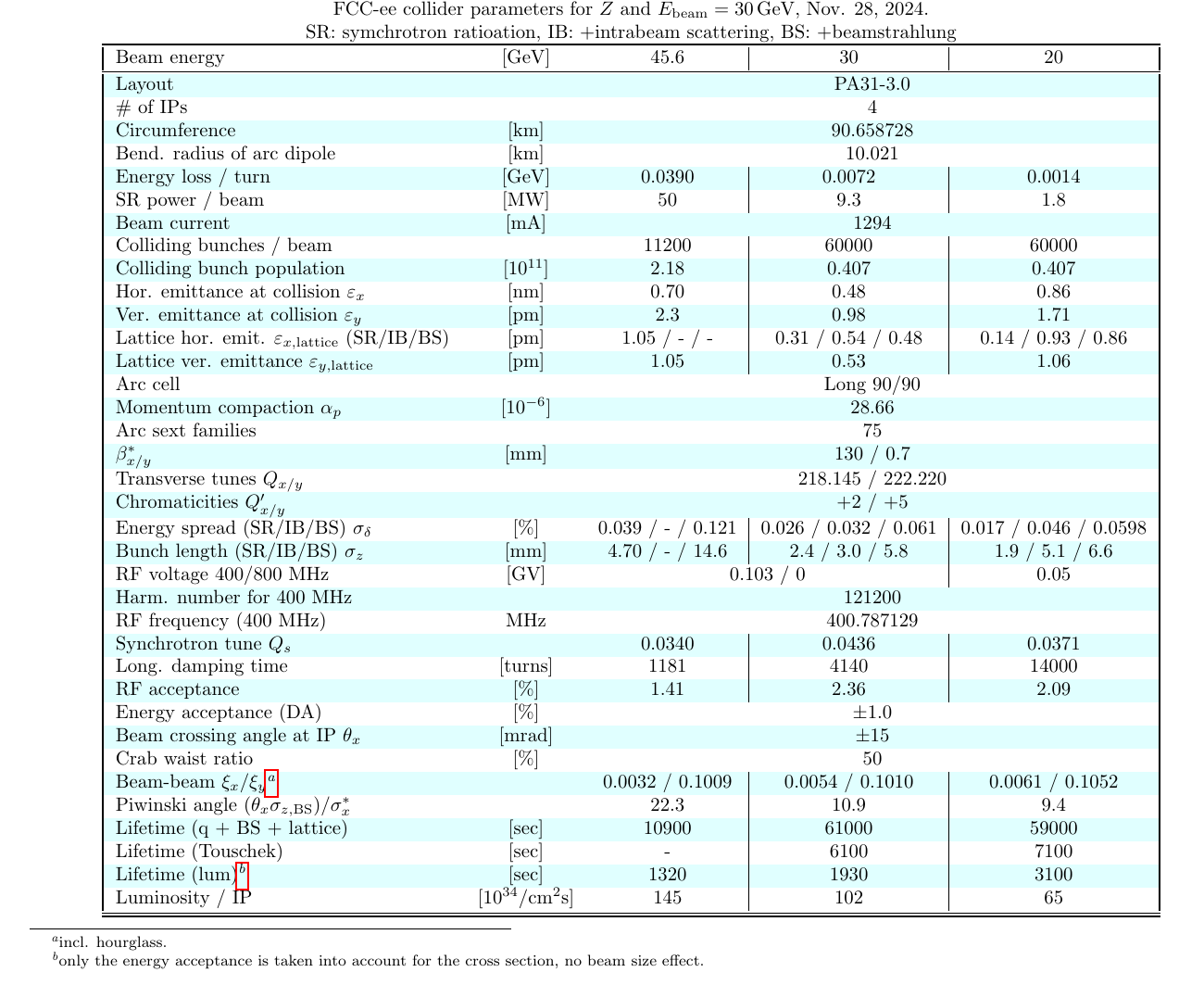}}
\label{tab:FCCee_low_sqrts}
\end{tabular}
\end{table}

\section{Summary}

Measurements of hadronic final states (HFS) in $\epem$ collisions at centre-of-mass (CM) energies between those achieved at B factories ($\sqrt{s} \approx 10\,\GeV$) and the Z pole ($\sqrt{s} \approx 91\,\GeV$) are scarce and subject to large experimental uncertainties. HFS in such an energy range can be exploited to perform multiple precision QCD measurements in the very clean $\epem$ collision environment. These involve the study of light-, heavy-quark and gluon jets properties, event shapes and fragmentation functions, valuable to advance our understanding of nonperturbative QCD as well as the theoretical description of heavy-quark mass effects.

A future collider such as FCC-ee offers two complementary options to access $\epem$ collisions at hadronic CM energies $\sqrt{s_\text{had}} \approx 20\mbox{--}80\,\GeV$, either exploiting initial- or final-state radiation (ISR/FSR) events during the very high-luminosity Z-pole run, and/or in dedicated short (about 1-month long) $\epem$ runs at $\sqrt{s} \approx 40\,\GeV$ and $60\,\GeV$. Using past ISR/FSR studies at LEP as a reference, we estimate that data samples with $\mathcal{O}(10^{9})$ hadronic events can be collected at the FCC-ee at different low CM energy points. Fast simulation studies for $\epem\to\qqbar$ collisions and different background processes, at $\sqrt{s} = 91.2\,\GeV$, smeared with the IDEA detector response and implementing the ISR/FSR selection criteria used at LEP, indicate that it is possible to obtain such data samples with reasonable purity for each $5$-$\GeV$ bin in effective CM energy over $\sqrt{s_\text{had}} = 20$--$80\,\GeV$. More elaborate studies, including simulations for different detectors, can be performed but we trust that such main conclusion will remain true. The technical feasibility of operating the FCC-ee at CM energies below the Z pole has not been studied, but simulations of the accelerator settings, starting off from the beam parameters for the Z-pole run without modifications of the machine, indicate that $\mathcal{O}(10^9)$ HFS data samples can also be collected within about 1-month of operation at the lowest energy achievable by the machine ($\sqrt{s}\approx 40\,\GeV$) and at an intermediate energy point ($\sqrt{s}\approx 60\,\GeV$).
The data collected through both approaches is of key importance for the physics program at the FCC-ee. Whereas the collection of the ISR/FSR events is already implicit in the baseline FCC-ee program, that from dedicated 1-month-long runs will have a superior precision and purity, but require specific machine and detector studies.

\paragraph*{Acknowledgments.} We are grateful to Patrick~Janot, Katsunobo~Oide, and Frank~Zimmermann for informative discussions, and to Patrick~Janot for constructive comments on the note. PM is funded by the European Union (ERC, grant agreement No. 101044599), and the work of PS is supported by ARC grant DP230103014. Views and opinions expressed are however those of the authors only and do not necessarily reflect those of the European Union or the European Research Council Executive Agency. Neither the European Union nor the granting authority can be held responsible for them.

\section*{References}
\printbibliography[heading=none]
\end{document}